\documentclass[11pt,a4paper]{article}
\usepackage[T1]{fontenc}

\usepackage[a4paper,top=2cm,bottom=2cm,left=3cm,right=3cm,marginparwidth=1.75cm]{geometry}
 
\usepackage{amsmath}
\usepackage{graphicx}
\usepackage{hyperref}

\newcommand{\keepcomment}{1} 
\usepackage[normalem]{ulem}
\ifnum\keepcomment=1
	\usepackage[colorinlistoftodos,textsize=scriptsize]{todonotes} 
    \newcommand{\stkout}[1]{\ifmmode\text{\sout{\ensuremath{#1}}}\else\sout{#1}\fi}
    
\else
	
	\usepackage[disable]{todonotes} 
\fi

\usepackage{amsfonts} 

\usepackage{calrsfs} 
\DeclareMathAlphabet{\pazocal}{OMS}{zplm}{m}{n}

\newcommand{\gls}[1]{\MakeUppercase{#1}}

\usepackage[affil-it]{authblk}

\makeatletter
\renewcommand{\section}{\@startsection {section}{1}{\z@}%
              {24pt}{12pt} {\large\scshape\bfseries}}

\renewcommand{\subsection}{\@startsection {subsection}{2}{\z@}%
             {12pt}{12pt}  {\itshape\bfseries}}

\usepackage{apacite}
\usepackage{natbib}

\title{\bfseries \normalsize On the Computation of Accessibility Provided by Shared Mobility }

\author[1]{Severin Diepolder*}

\author[2]{Andrea Araldo}
\author[3]{Tarek Chouaki}
\author[4]{Santa Maiti}
\author[5]{Sebastian Hörl}
\author[6]{Constantinos Antoniou}

\affil[1]{Student, Technical University of Munich, Germany}
\affil[2]{Associate Professor, SAMOVAR, Télécom SudParis, Institut Polytechnique de Paris, France}
\affil[3]{PhD Candidate, SystemX, France}
\affil[4]{Postdoc, Technical University of Munich, Germany }
\affil[5]{PhD, SystemX, France}
\affil[6]{Professor, Technical University of Munich, Germany}

\date{\vspace{-5ex}}

\begin{document}
\maketitle

\section*{Short summary}\small
Shared Mobility Services (SMS), e.g., Demand-Responsive Transit (DRT) or ride-sharing, can improve mobility in low-density areas, often poorly served by conventional Public Transport (PT). Such improvement is mostly quantified via basic performance indicators, like wait or travel time. However, accessibility indicators, measuring the ease of reaching surrounding opportunities (e.g., jobs, schools, shops, ...), would be a more comprehensive indicator. To date, no method exists to quantify the accessibility of SMS based on empirical measurements. Indeed, accessibility is generally computed on graph representations of PT networks, but SMS are dynamic and do not follow a predefined network. We propose a spatial-temporal statistical method that takes as input observed trips of a SMS acting as a feeder for PT and summarized such trips in a graph. On such a graph, we compute classic accessibility indicators. We apply our method to a MATSim simulation study concerning DRT in Paris-Saclay.


\textbf{Keywords}: Accessibility; Public Transport; Shared Mobility;

\section{Introduction}
Location-based accessibility measures the ease of reaching surrounding opportunities via transport (\cite{miller2020accessibility}). Accessibility provided by conventional PT is generally poor in low-demand areas, e.g., suburbs (\cite{badeanlou2022ptanalysistool}), because a high frequency and high coverage service in such areas would imply an unaffordable cost per passenger. Poor PT accessibility in the suburbs makes them car-dependent, which prevents urban regions from being sustainable (\cite[Section~2.2]{Boussauw2022}).
SMS, e.g., Demand-Responsive Transit (DRT), ride-sharing, carpooling, car-sharing, are potentially more efficient than conventional PT in the suburbs (\cite{Calabro2021}). However, their current deployment is commonly led by private companies targeting profit maximization. This may turn SMS into additional source of congestion and pollution (\cite{Marshall2019, Erhardt2019a}). 

We believe that SMS deployment should be overseen by transport authorities under the logic of accessibility improvement. To this aim, a method is needed, able to compute impact of SMS on accessibility, based on empirically observed trips. To the best of our knowledge, this paper is the first to propose such a method. \cite{chandra2013accessibility} study how DRT improves connection to conventioal PT stops, without considering the impact on accessing opportunities.
%
%
\cite{nahmias2021drtaccessibility} and \cite{zhou2021simulating} calculate accessibility from Autonomous Mobility on Demand, based on utilities perceived by agents within simulation. By contrast, our method computes accessibility solely based on observed SMS trip times, either from the real world or simulation. A first attempt of integrating SMS into the graph-based description of PT is done by \cite{hasif2022graph}. However, they use analytic models to model SMS performance and thus fail to give real insights adapted to the areas under study. Our effort consists instead of estimating accessibility from empirical observations via spatial-temporal statistics.
General Transit Feed System (GTFS) is the standard data format for PT schedules. Recently, the GTFS-Flex extension allows also describing 
SMS (\cite{craig2020gtfs}). Although uur estimates could thus be fed into GTFS-Flex data, for the sake of simplicity, we use plain GTFS instead.

Our contribution consists in developing a spatial-temporal statistical pipeline to transform SMS trip data observations in a graph representation, on top of which well-established accessibility computation can be performed. 
The observations that can be taken as input might come from real measurements or from simulation. This paper's
observations come from a MATSim simulation study of DRT deployment in Paris Saclay, from \cite{Chouaki2023}.

By providing a first method to compute the accessibility of SMS on empirical observations, this work can contribute to a better understanding of the potential of SMS and guide their future deployment.

\section{Methodology}
\label{sec:methodology}

\subsection{Accessibility}
As in (\cite{biazzo2019accessibility}), the study area is tessellated in hexagons with a grid step of 1km, whose centers $\mathbf{u}\in\mathbb{R}^2$ are called centroids and denoted with set $\pazocal{C} \subseteq   \mathbb{R}^2$. Each hexagon contains a certain quantity of \emph{opportunities}, e.g., jobs, places at school, people. With $O_\mathbf{u}$ we denote the opportunities in the hexagon around $\mathbf{u}$ and with $T(\mathbf{u},\mathbf{u}', t)$ the time it takes to arrive in $\mathbf{u}'$, when departing from $\mathbf{u}$ at time $t$.
As in \cite{miller2020accessibility}, accessibility is the amount of opportunities that one can reach departing from $\mathbf{u}$ at time of day $t$ within time $\tau$:
\begin{align}
    acc(\mathbf{u}) \equiv \sum_{\mathbf{u}'\in\pazocal{C}(\mathbf{u},t)} O_{\mathbf{u}'}.
    \label{eq:acc}
\end{align}
$\pazocal{C}(\mathbf{u},t)=\{\mathbf{u}\in\pazocal{C} | T(\mathbf{u},\mathbf{u}', t)\le \tau\}$ is the set of centroids reachable within $\tau$. By improving \gls{pt}, such set can be enlarged such as to consent to reach more opportunities. In this work, the opportunities are the number of people (residents) that can be reached.
$T(\mathbf{u},\mathbf{u}', t)$ is always computed on a graph representation of the transport network. However, SMS are not based on any network. Our effort is thus to build a graph representation of SMS, despite the absence of a network model.

\subsection{Time-Expanded Graph Model of conventional \gls{pt}}
\label{time-expanded graphs for pt}

\begin{figure}
\center \includegraphics[width=100mm]{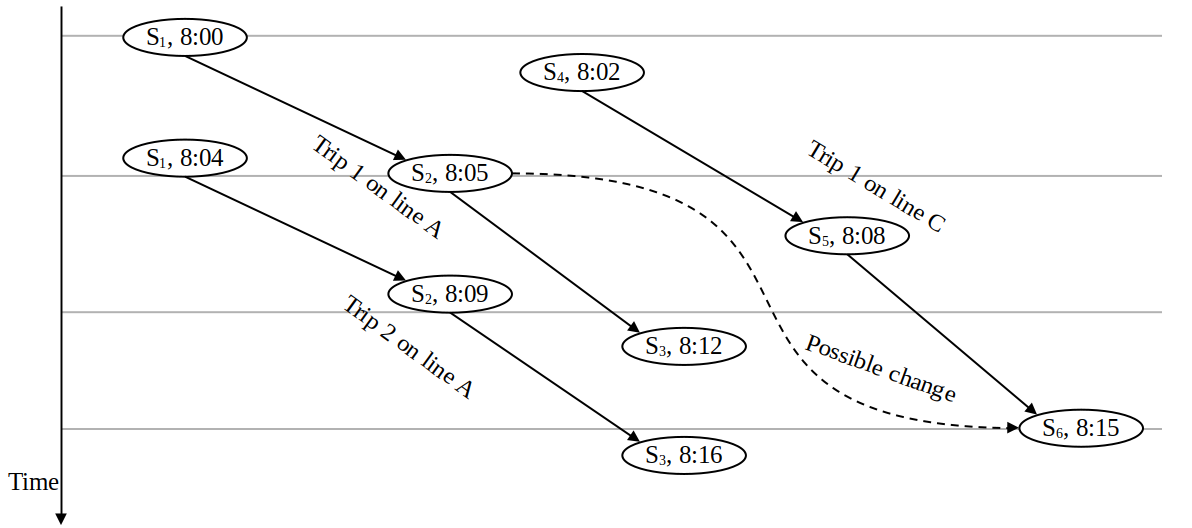}
\caption[time-expanded Graph for Dynamic Feeder Service]{Time-expanded graph, representing two trips on line A and one trip on line B, as well as a potential change.}
\label{fig:graph}
\end{figure}

Inspired by~\cite{fortin2016innovative} and~\cite{hasif2022graph}, we model \gls{pt} as a time-expanded graph~$\pazocal{G}$, compatible with the GTFS format. The nodes of $\pazocal{G}$ are \emph{stoptimes}. Stoptime $(\mathbf{s},t)$ indicates the arrival of a \gls{pt} vehicle at stop $\mathbf{s}\in\mathbb{R}^2$ (modeled as a point in the plane) at time $t\in\mathbb{R}$. Different trips on a certain line are represented as sequences of different stoptimes, as in Figure~\ref{fig:graph}, as well as potential line change,  within 15 minutes walk, assuming 5 Km/h walk speed, if it is possible to arrive at the new line on time. When a user departs at time $t_0$ from location $\mathbf{x}$ for location $\mathbf{x}'$, they can simply walk (but no more than the maximum walk time). Or they can walk to $\mathbf{s}$, board a PT vehicle at $t$ (corresponding to a stoptime $(\mathbf{s},t)$, use \gls{pt} up to a stoptime $(\mathbf{s}',t')$ and from there walk to $\mathbf{x}'$. The arrival time
at $\mathbf{x}'$ will be $t'$ plus the time for walking.
Users are assumed to always choose the path with the earliest arrival time. Path computation is performed within CityChrone (\cite{biazzo2019accessibility}). No capacity constraints are considered.

\subsection{Integration of shared mobility into the time-expanded graph} \label{time-expanded graphs for shared mobility}

SMS is assumed to provide a feeder service to traditional \gls{pt}. In a feeder area $\pazocal{F}(\mathbf{s})\subseteq \mathbb{R}^2$ around some selected stops $\mathbf{s}$ (which we also call \emph{hubs}), SMS provide connection to and from $\mathbf{s}$. The set of centroids in such an area is $\pazocal{C}(\mathbf{s})=\pazocal{C}\cap \pazocal{F}(\mathbf{s})$.
In this section, we will focus on access trips (from a location to a PT stop) performed via SMS. The same reasoning \textcolor{blue}{applies to} 
egress trips, \textit{mutatis mutandis}.
We assume to have a set $\pazocal{O}$ of observations. Each observation $i\in\pazocal{O}$ corresponds to an access trip and contains:
\begin{itemize}
    \item Time of day $t_i\in \mathbb{R}$ when the user requested a trip to the flexible service
    \label{line:information}
    \item Location $\mathbf{x}_i\in\mathbb{R}^2$ where the user is at time $t_i$
    \item Station $s_i$ where the user wants to arrive via the SMS feeder service
    \item Duration $w_i$ indicating the wait time before the user is served: it can be the time passed between the time of request and the time of pickup from a vehicle, in case of ride-sharing, \gls{drt} or carpooling; it can be the time to wait until a vehicle is available at the docks in a car-sharing or bike-sharing system.
    \item Travel time $y_i$: time spent in the SMS vehicle to arrive at $\mathbf{s}$.
\end{itemize}


We interpret $y_i$ and $w_i$ as realizations of spatial-temporal random fields (\cite{Handcock1994}): for any time of day $t\in\mathbb{R}$ and physical location $\mathbf{x}\in\pazocal{F}(\mathbf{s})$, random variables $W^\mathbf{s}(\mathbf{x},t), Y^\mathbf{s}(\mathbf{x},t)$ represent the times experienced by a user appearing in $t$ and $\mathbf{x}$, for any stop $\mathbf{s}$. In the following subsection we will compute estimations $\hat w^\mathbf{s}(\mathbf{u},t), \hat y^\mathbf{s}(\mathbf{u},t)$ of expected values $\mathbb E[W^\mathbf{s}(\mathbf{u},t)],\mathbb E[Y^\mathbf{s}(\mathbf{u},t)]$ at centroids $\mathbf{u}\in\pazocal{C}(\mathbf{s})$.

\begin{figure}
    \centering
    \includegraphics[width=0.35\textwidth]{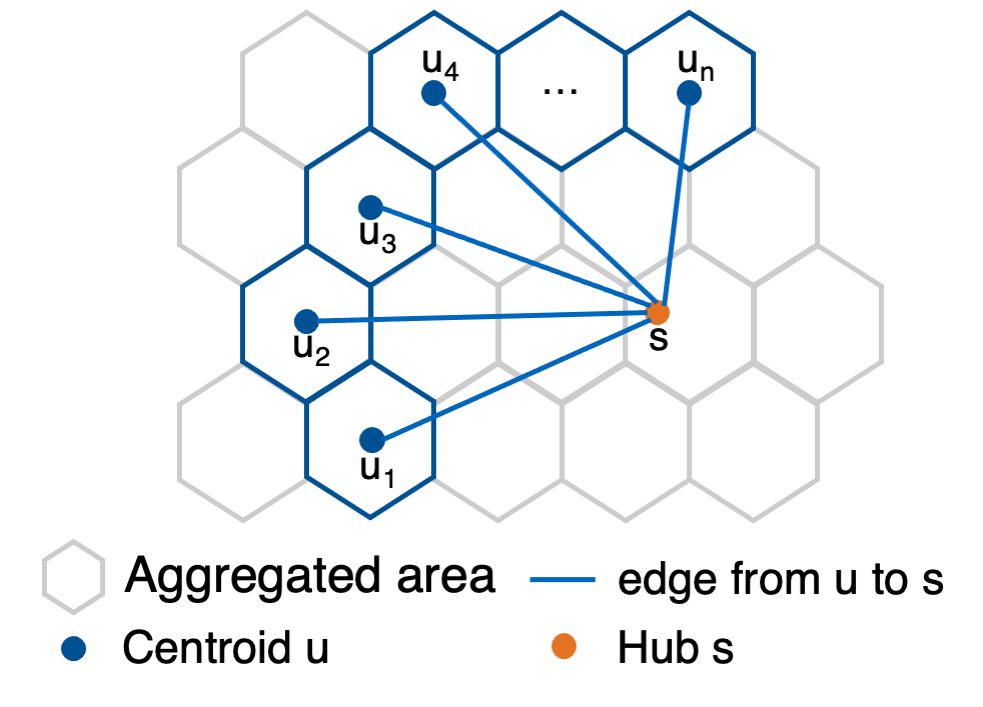}
    \caption{Hub and virtual trips provided by SMS}
    \label{fig:hub}
\end{figure}

To integrate SMS into \gls{pt} graph $\pazocal{G}$, SMS are represented as a set of ``virtual'' trips, running between centroid $\mathbf{u}\in\pazocal{C}(\mathbf{s})$ and hub $\mathbf{s}$ (Figure~\ref{fig:hub}).
Each trip has travel time $\hat y^\mathbf{s}(\mathbf{u},t)$. The access connection between centroid $\mathbf{u}$ and hub $\mathbf{s}$ is modeled as a sequence of trips, corresponding to stop times $(\mathbf{u}, t_j)$, for different values of departure time. We thus have to compute the list of such departure times. To do so, we interpret the inter-departure time between sich trips as a random field $H^\mathbf{s}(\mathbf{x},t)$, which represent a ``virtual'' headway. The value of such an interval in $\mathbf{x}$ and $t$ is also a spatial-temporal random field. We use the common approximation ((2.4.28) from~\cite{cascetta2009transportation} and related assumptions): $H^\mathbf{s}(\mathbf{x},t)=2\cdot ^\mathbf{s}(\mathbf{x},t)$.
Therefore, we separate stoptimes by $2\cdot w^\mathbf{s}(\mathbf{u},t)$. More precisely, the stoptimes corresponding to access trips departing from centroid $\mathbf{u}$ to hub $\mathbf{s}$ are:
\begin{align}
    (\mathbf{u},t_0),
    \nonumber
    \\
    (\mathbf{u},t_j)
    && \text{where }t_j= t_{j-1}+ 2\cdot \hat w^\mathbf{s}(\mathbf{u},t_{j-1})
    & \text{for }j=1,2,& \text{ until 11:59 pm},
    \label{eq:virtual_stoptimes}
    \\
    (\mathbf{u},t_j) 
    && \text{where }t_j = t_{j+1}-2\cdot \hat w^\mathbf{s}(\mathbf{u},t_{j+1})
    & \text{for }j=-1,-2,& \text{ until 00:00 am}.
    \nonumber
\end{align}

Correspondent stoptimes are added to represent the arrival of access trips $(\mathbf{s},t_j+\hat y^\mathbf{s}(\mathbf{u},t_j))$ and an edge between each departure stoptime and the respective arrival stoptime is added. A similar process is applied for egress trips. At the end of the described process, time-expanded graph $\pazocal{G}$ is enriched with stoptimes and edges representing SMS trips. Having done so, it is possible to reuse accessibility calculation methods for time-expanded graphs, such as CityChrone~\cite{biazzo2019accessibility}, with no modifications required. 

\subsection{Estimation of Waiting and Travel Times} \label{m_scarce_data}

We now explain how we construct estimation $\hat w^\mathbf{s}(\mathbf{u},t)$ used in the previous subsection, for access SMS trips only. Similar reasoning can be applied to $\hat y^\mathbf{s}(\mathbf{u},t)$ and egress trips. We assume random field $W^\mathbf{s}(\mathbf{x},t)$ is approximately temporally stationary within each timeslot:
\begin{flalign}
    && W^\mathbf{s}(\mathbf{x},t)=W^\mathbf{s}(\mathbf{x},t_k),
    &&
    \forall \mathbf{x}\in\mathbb{R}^2, \forall t\in[t_k, t_{k+1}[, \forall \text{ station }\mathbf{s}
    &
    \label{eq:stationary}
\end{flalign}

For any timeslot, we thus just need to find estimation $\hat w_{t_k}^\mathbf{s}(\mathbf{u})$ of the expected values of random field $W_{t_k}^\mathbf{s}(\mathbf{x}) \equiv W^\mathbf{s}(\mathbf{x},t_k)$.
First the observations $\pazocal{O}$ are projected onto time-slot $[t_k, t_{k+1}]$:
\begin{equation}
        \pazocal{O}_{t_k}^\mathbf{s}
        \ \ \ \equiv
        \left\{
            \text{observation }i=(\mathbf{x}_i,w_i,y_i) | i\in\pazocal{O}, t\in[t_k,t_{k+1}[,
            i \text{ is related to an access trip to }\mathbf{s}
        \right\}
\end{equation}
Estimation $\hat w_{t_k}^\mathbf{s}(\mathbf{u})$ is computed by Ordinary Kriging (~\cite{Geosciences}) on the observations $\pazocal{O}_{t_k}^\mathbf{s}$ as a convex combination of observations $w_i$:
\begin{equation}
    \hat w_{t_k}^\mathbf{s}(\mathbf{x}) = \sum_{i\in\pazocal{O}_{t_k}^\mathbf{s}} \lambda_i\cdot w_i
    \label{eq:kriging-estimation}
\end{equation}
    
In short (details can be found in Section~19.4 of \cite{Chilès2018ok}), coefficients $\lambda_i$ are computed based on a \emph{semivariogram} function $\gamma_{t_k}^\mathbf{s}(d)$, which obtained as a linear regression model, with predictors $d_{i,j}$ (distances between all pair of observations) and labels $\gamma_{i,j}$, which are called \emph{experimental seminariances}:
\begin{equation}
    \gamma_{i,j} \equiv \frac{1}{2}\cdot (w_i-d_j)^2
    \label{eq:gamma_ij}
\end{equation}
    
 The underlying assumption here is that correlation between wait times in different locations vanishes with the distance between such locations. The semivariogram gives the ``shape'' of this vanishing slope. In estimation~\eqref{eq:kriging-estimation}, closer observations will have a higher weight. Under hypothesis on spatial stationarity and uniformity in all directions (\cite{Columbia}), Theorem~2.3 of \cite{Szidarovsky1985}\label{line:unbiased} proves that Kriging gives an aymptotically biased estimator: as the number of observations tends to infinite, $\hat w_{t_k}^\mathbf{s}(\mathbf{x})$ tends to the ``true'' $\mathbb{E}[W_{t_k}^\mathbf{s}(\mathbf{x})]$.

Note that, by means of interpolation on a limited set of observed trips, the method described here is meant to infer the potential to access opportunities, also via trips that may not have observed yet.

\section{Implementation}
\begin{figure}
    \noindent\hspace{0.5mm}\includegraphics[width=149mm]{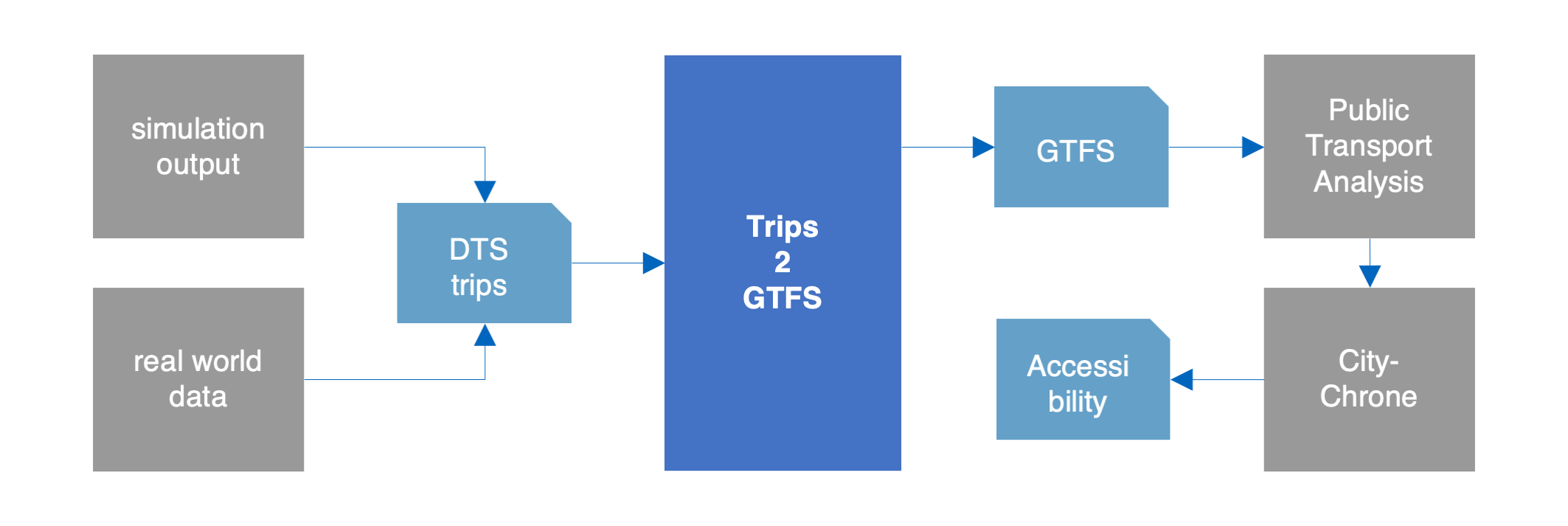}
    \caption[Schema of Implementation for DRT Services]{Implementation pipeline. CityChrone is from \cite{biazzo2019accessibility}.}
    \label{fig:workflow}
\end{figure}

    The methodology of Section~\ref{sec:methodology} is implemented in a Python pipeline, which we release as open source (\cite{githubrepo}) and is depicted in Figure~\ref{fig:workflow}. 
    \begin{enumerate}
        \item We first get centroids and cells performing the tessellation via CityChrone.
        \item We read the file containing the observations (SMS trips). Such a file can be a simulation output or measurements of real SMS. Each observation includes the same information as in page~\pageref{line:information}. Observations are stored in a dataframe.
        \item We assume SMS is deployed as feeder (as it is the case for the MATSim simulation on which we perform our analysis). Therefore, we can classify every SMS trip as either access or egress, depending on whether the origin or the destination is a PT stop.
        \item To establish the feeder area $\pazocal{F}(\mathbf{s})$, we find among the observations $\pazocal{O}$ the furthest cell from $s$ in which a trip to/from $\mathbf{s}$ has occurred. All cells within such a distance, are assumed to be in $\pazocal{F}(\mathbf{s})$. Observe that feeder areas of different hubs may overlap.
        \item We group observations in timeslots (Figure~\ref{fig:distribution_wt_access}).
        \item In each time slot $[t_k,t_{k+1}[$ and each centroid $\mathbf{u}$ around each stop $\mathbf{s}$, we perform Kriging via library $\texttt{pyInterpolate}$ (\cite{molinski2022pyinterpolate}) to obtain estimations $\hat w^\mathbf{s}(\mathbf{u})$ and $\hat y_{t_k}^\mathbf{s}(\mathbf{u})$.
        \item We obtain stoptimes and edges using the estimations above, as specified in ~\eqref{eq:virtual_stoptimes}. We add stoptimes and edges to the GTFS data of conventional PT, following the specifications in \cite{gtfsReference}.
        \item We give the obtained graph to CityChrone, which will give us accessibility scores in all the centroids.
    \end{enumerate}

\section{Results and discussion}

\begin{table}
   
  \begin{center}
    \caption{Parameters used for the numerical results.}
     \vspace{2ex}
    \label{tab:parameters}
    \begin{tabular}{l|c|r} 
      \textbf{Parameter} & \textbf{Value} & \textbf{Reference}\\
      \hline
      Side of a hexagon (tessellation) & 1km & \cite{badeanlou2022ptanalysistool}\\
      $\tau$ (Equation~\eqref{eq:acc}) & 1 hour & \cite{badeanlou2022ptanalysistool}\\
      Total number of DRT trips & 14700 & \\
        - access trips & 5289 & \\
        - egress trips & 9412 & \\
    Total number of hubs & 16 &\\
    Walk times & & Computed via OpenStreetMap\\
    Population Distribution & & from the simulation scenario from \cite{Chouaki2023}
      
    \end{tabular}
  \end{center}
\end{table}

\subsection{Data Source of the observations} \label{Scenariodescribtion}

\begin{figure}      
\noindent\hspace{0.5mm}\includegraphics[width=148mm]{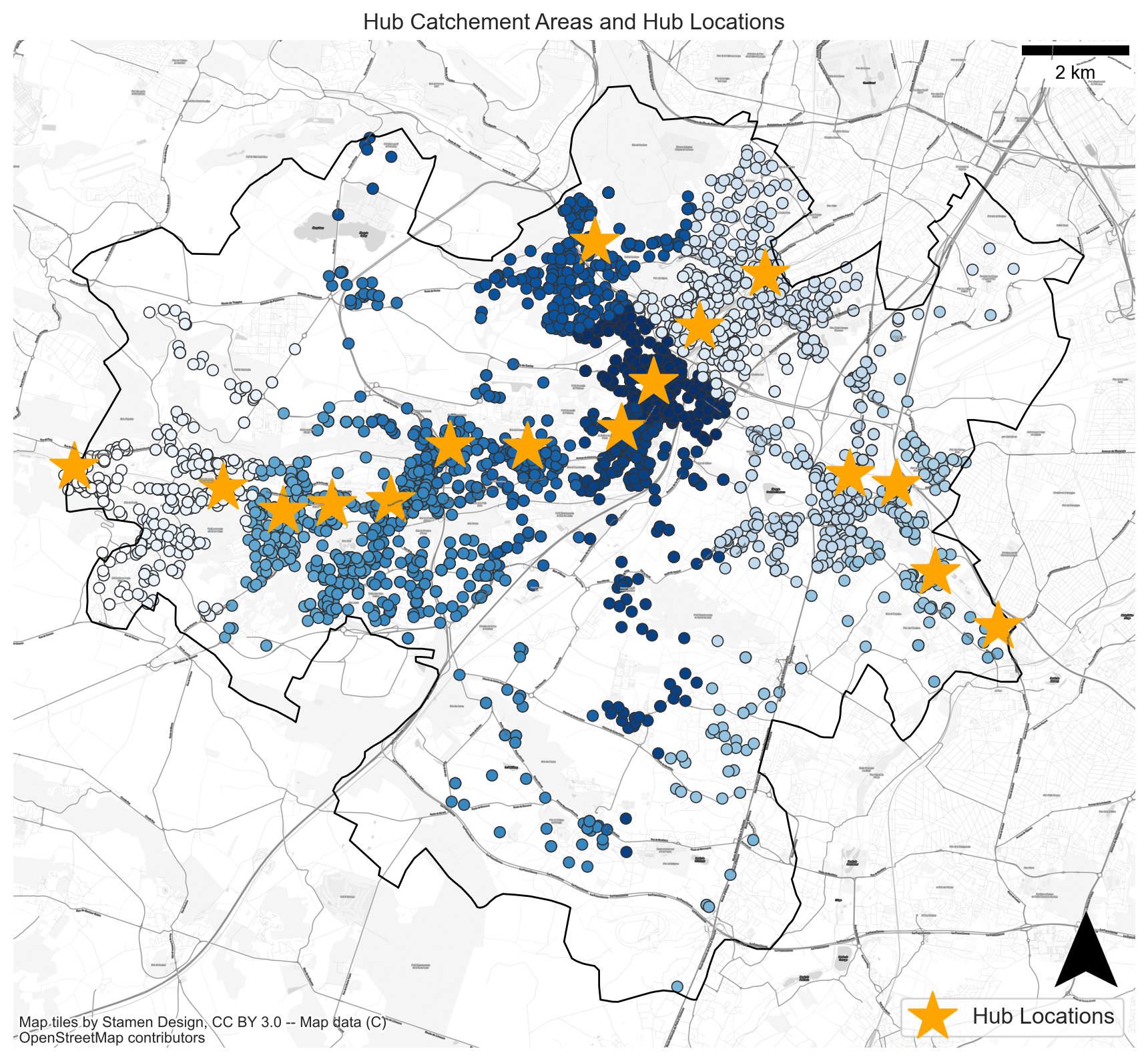}
\caption[Hub Catchment Areas and Hub Locations]{ Hub Catchment Areas and Hub Locations. Each dot corresponds to the origin of one trip observed during the simulation. The differentiation in color of the observed trip origins indicates the catchment by different hubs}
\label{map:drt_hubs}
\end{figure}

The observation dataset in this study comes from a MATSim simulation, from \cite{chouaki2021agent,Chouaki2023}, of door-to-door Demand-Responsive Transit (DRT), deployed as a feeder to and from conventional PT, in Paris-Saclay.
The area in which DRT is deployed is depicted in Figure~\ref{map:drt_hubs}, but the entire Paris Region is simulated. Scenario parameters are in Table~\ref{tab:parameters}.

\begin{figure}      
\center\includegraphics[width=100mm]{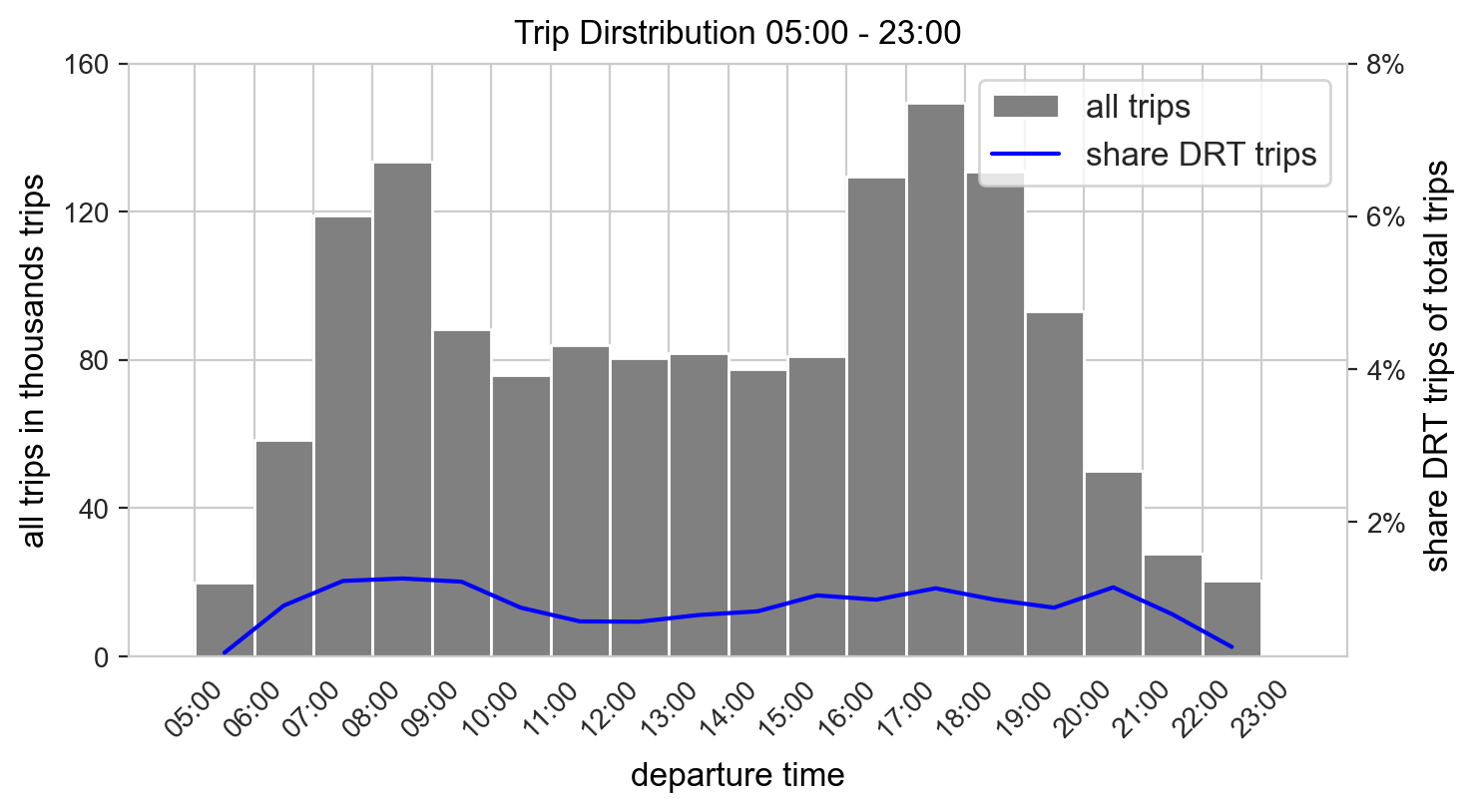}
\caption[Total Trips and DRT Trips over Time]{ Trips over time. One trip is defined by the departure within the study area of Paris Saclay. A trip consisting out of multiple legs (e.g. walk + drt + PT is considered as one trip)}
\label{fig:trips-tt-wt}
\end{figure}

\subsection{Analysis of Temporal and Spatial Patterns of DRT trips}
Figure~\ref{fig:trips-tt-wt} clearly shoes morning peak $[7:00, 10:00[$, evening peak $[16:00, 19:00[$ and off-peak (all the other intervals).
%
%

\begin{figure}      
\noindent\hspace{0.5mm}\includegraphics[width=148mm]{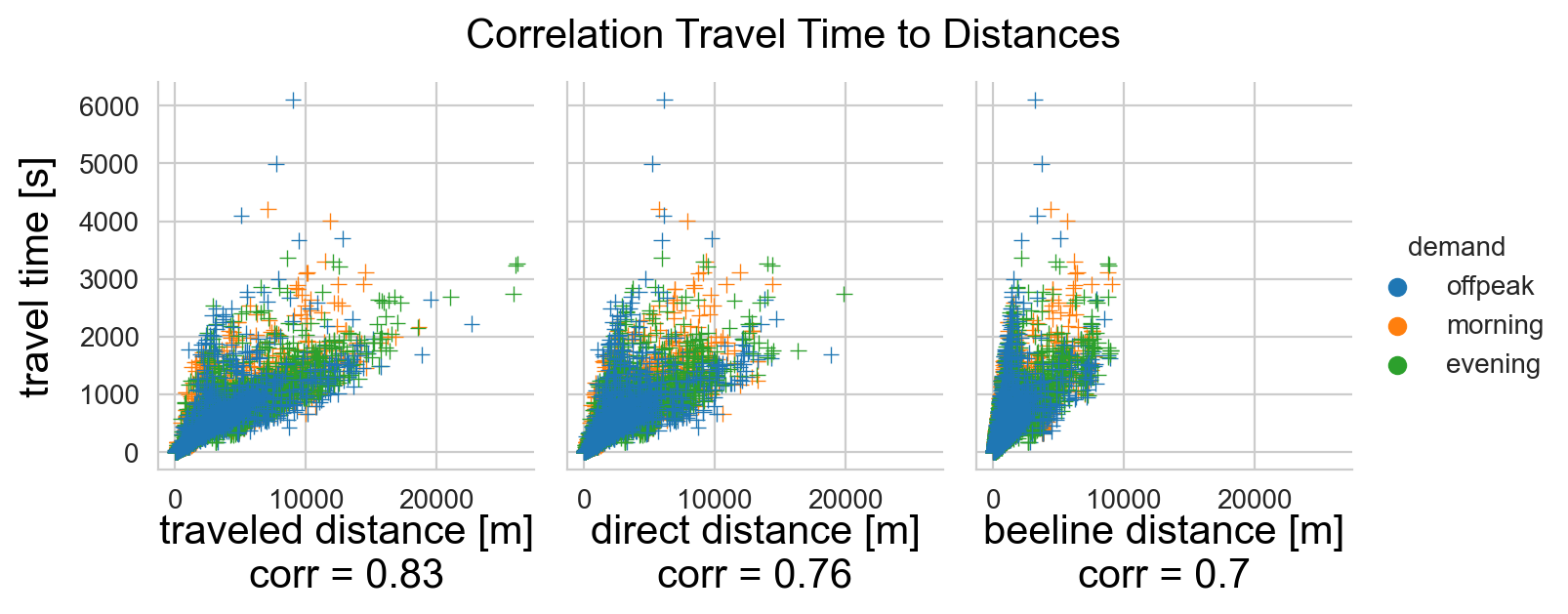}
\caption[Pair-plot Travel Time - Traveled Distance, Direct Distance, Beeline Distance]{Relation between travel time and distance measures. Traveled distance is the actual Km traveled by the user inside the DRT vehicle. Direct distance is the one from the shortest road network road from the origin centroid to the hub. Beeline is the Euclidean distance.}
\label{fig:corr_pairplot_tt}
\end{figure}
\begin{figure}      
\noindent\hspace{0.5mm}\includegraphics[width=148mm]{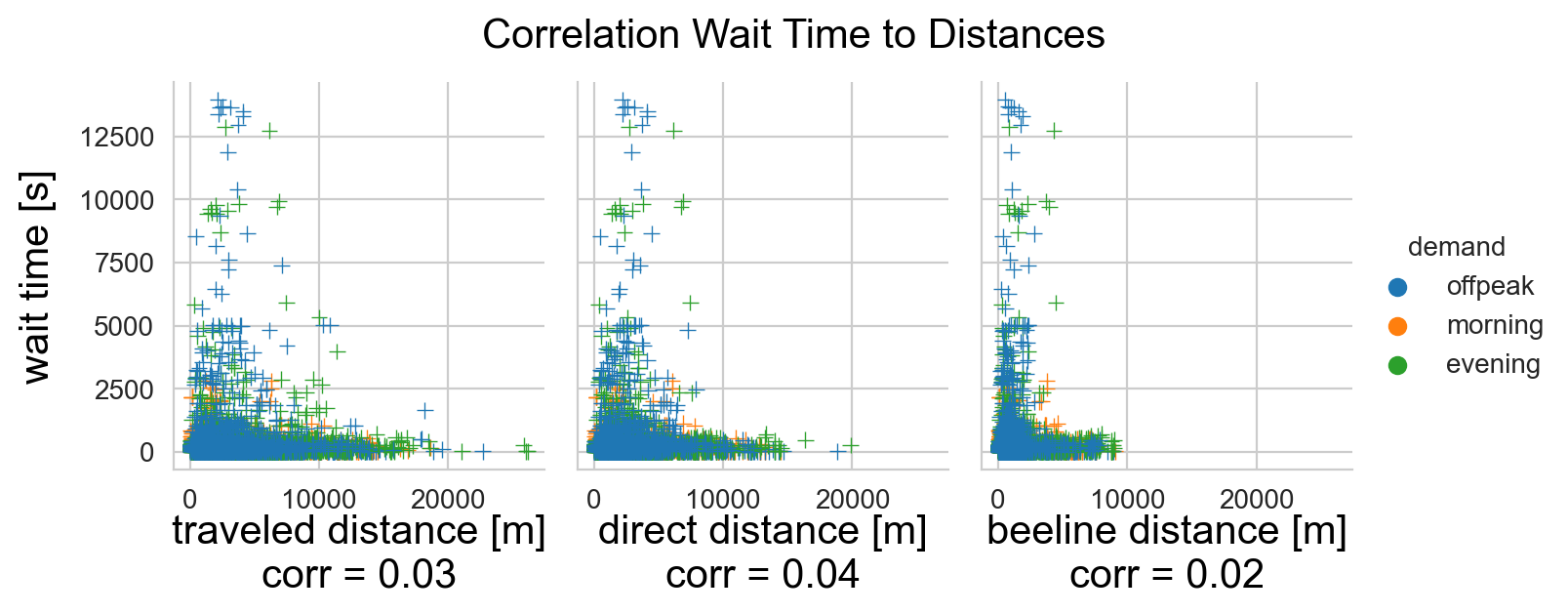}
\caption[Pair-plot Wait Time - Traveled Distance, Direct Distance, Beeline Distance]{Relation between wait time and distance measures.}
\label{fig:corr_pairplot_wt}
\end{figure}

The following figures concern DRT trips toward/from all hubs, without distinguishing between hubs.
Figure~\ref{fig:corr_pairplot_tt} is a negative result: travel times (figures on the right) do not appear to be spatially stationary (the distribution of values measured close to the related PT stops is different than further). Therefore, our estimations are not guaranteed to be asymptotically unbiased (page~\pageref{line:unbiased}). In our future work, we will explore indirect estimation of travel times through other indicators, e.g., the detour factor of DRT, which respect the requirements for the unbiasedness of Kriging. Correlation between wait times and distance is instead weaker (Figure~\ref{fig:corr_pairplot_wt}).

\begin{figure}      
\center\includegraphics[width=100mm]{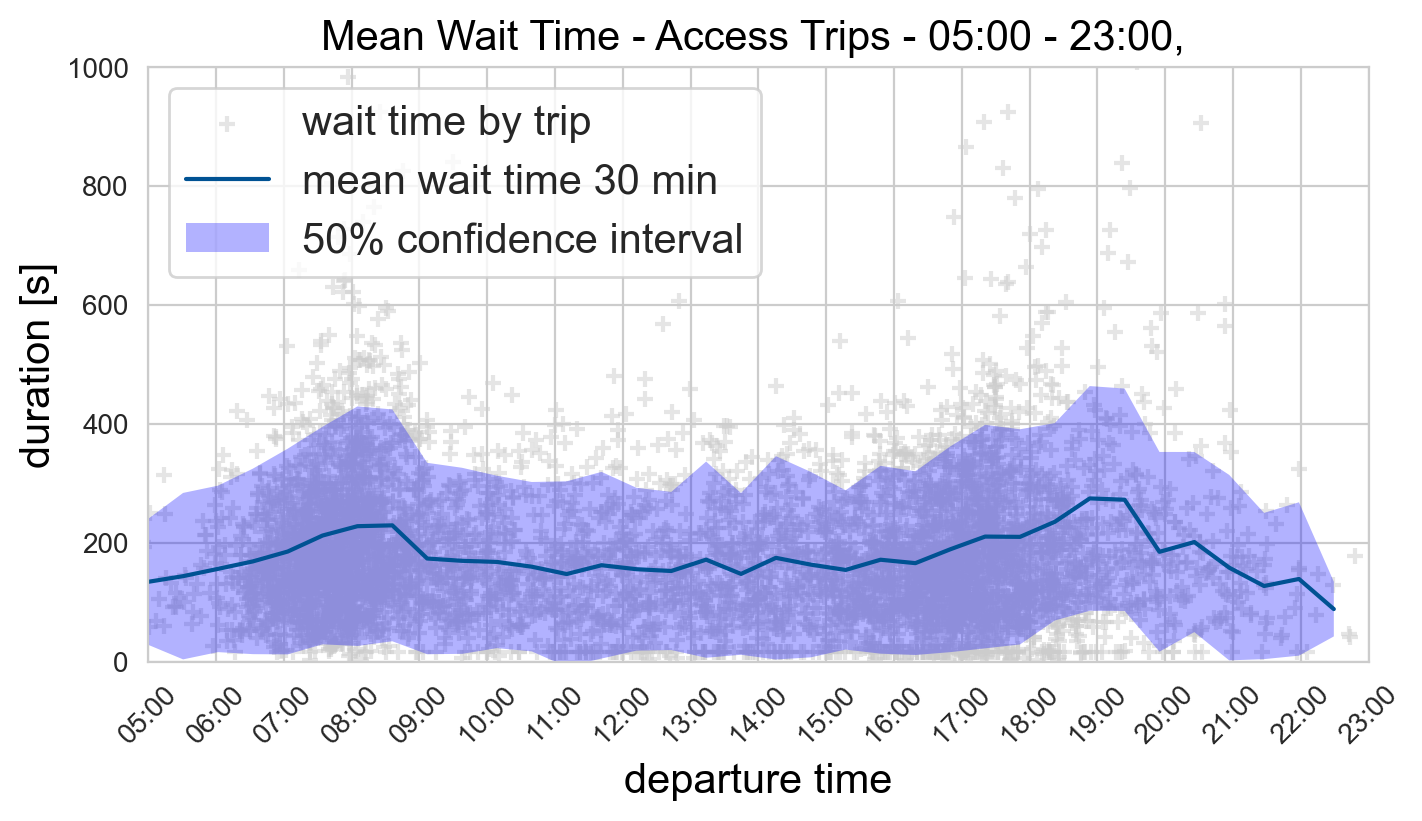}
\caption[Mean Wait Time]{Mean Wait Time - A moving average of wait time during one day}
\label{fig:wt_mean}
\end{figure}
Figure~\ref{fig:wt_mean} shows that wait time follows expected peak/off-peak patterns. Values are generally slow since the simulation is configured so that a DRT trip is accepted only if it the dispatcher predicts it is possible to serve it within 10 minutes. All wait times exceeding this limits might be due to the dispatcher not taking traffic correctly into account.

\subsection{Estimation of Waiting and Travel Times}
\label{sec:quality}

\begin{figure}      
\noindent\hspace{0.5mm}\includegraphics[width=148mm]{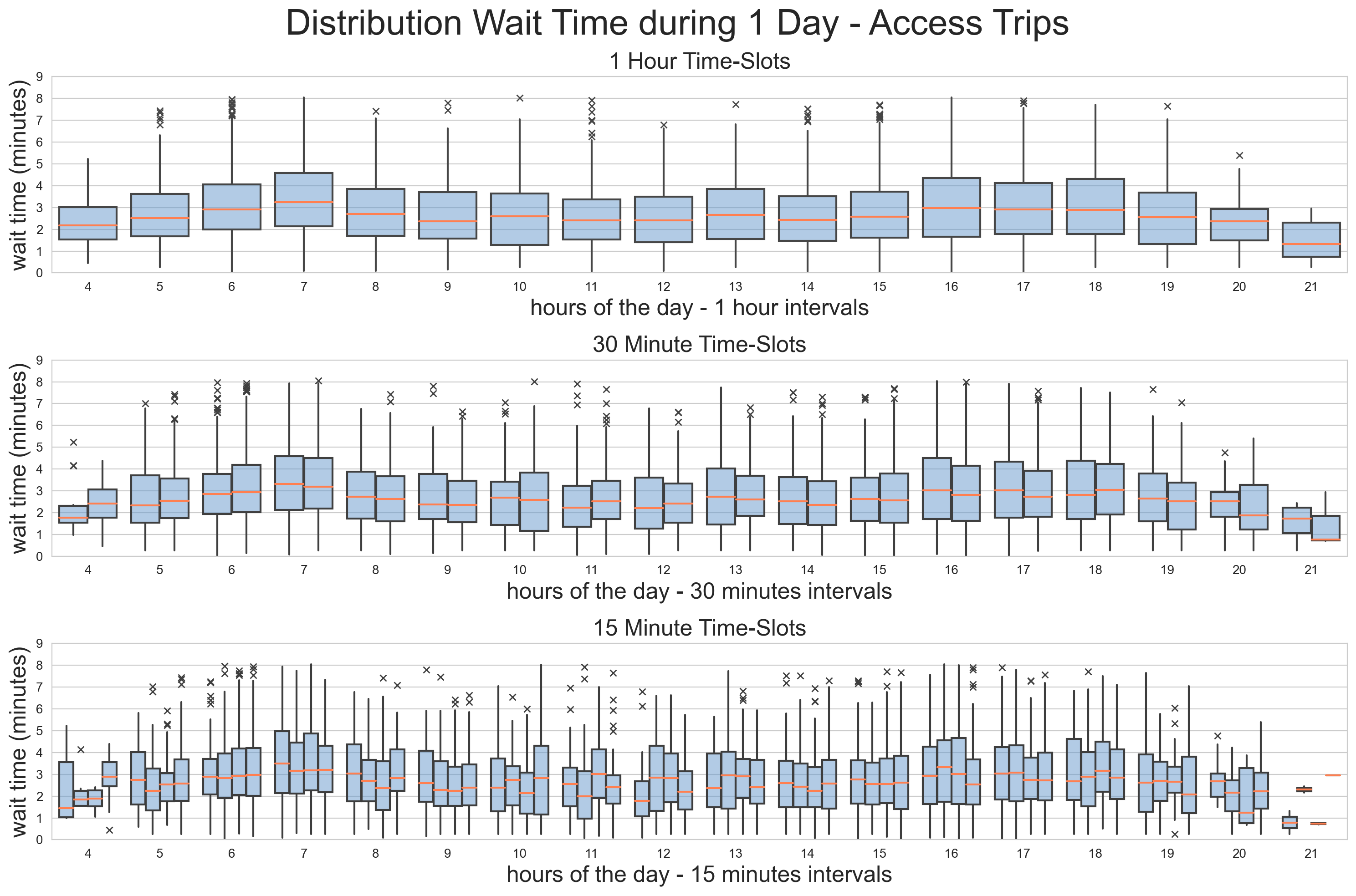}
\caption[Confidence and Preservation of Detail for Time Intervals]{Comparison of different timeslot sizes. Values exceeding 10 minutes are not depicted, as they are due to simulation events unpredictable for the SMS dispatcher}
\label{fig:distribution_wt_access}
\end{figure}
Figure~\ref{fig:distribution_wt_access} shows that timeslots of 1h preserve the temporal pattern of trips, so 1h should be preferred to smaller timeslots, so as to perform Kriging with as many observations as possible.

\begin{figure}      
\center\includegraphics[width=120mm]{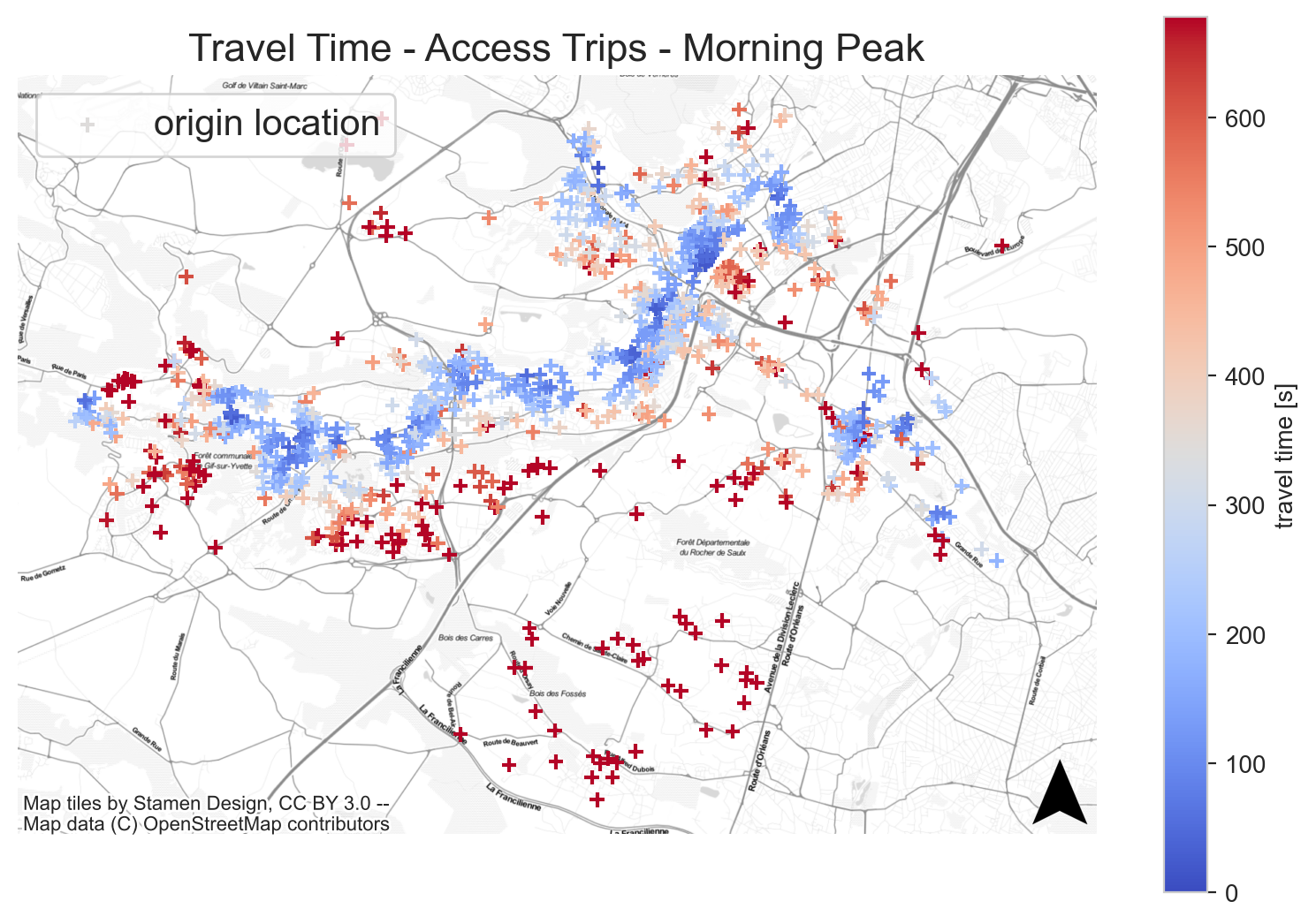}
\caption[Spatial Trend Travel Time]{Spatial Trend Travel Time for access time, morning peak (evening and off-peak show similar trends). }
\label{map:tt_space}
\end{figure}
\begin{figure}      
\center\includegraphics[width=100mm]{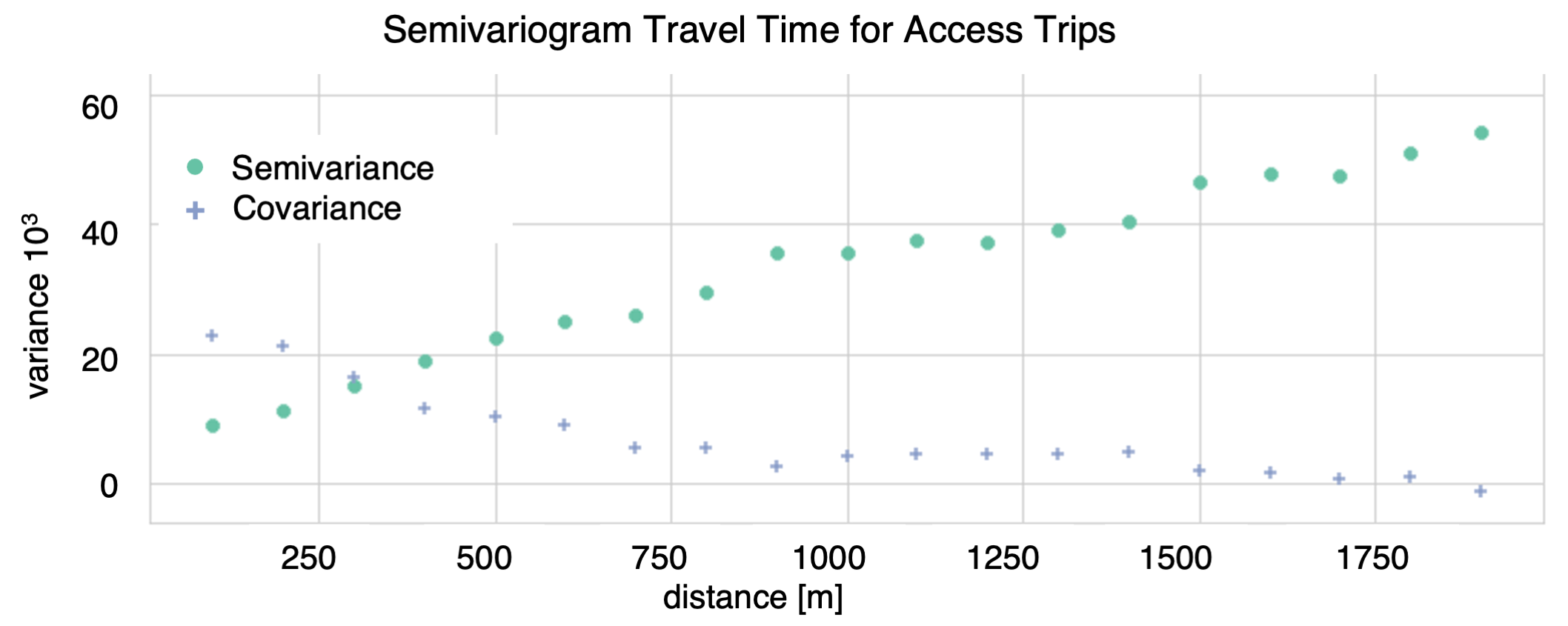}
\caption[Semivariogram Travel Time of Access Trip Observations]{Spatial correlation of travel time observations}
\label{semivar:tt}
\end{figure}
Within each timeslot, estimation of wait and travel times is based on Kriging, which exploits spatial correlation. First, we note in Figure~\ref{map:tt_space} that travel times close to hubs are shorter than further away. Then, we note that the experimental semivariance in Figure~\ref{semivar:tt}, i.e., the $\gamma_{i,j}$ between pair of observation $i,j$ (Equation~\eqref{eq:gamma_ij}), increases with the spatial distance between the observations: the closer the observations, the more similar are the respective travel times measured therein.

\begin{figure}      
\center\includegraphics[width=120mm]{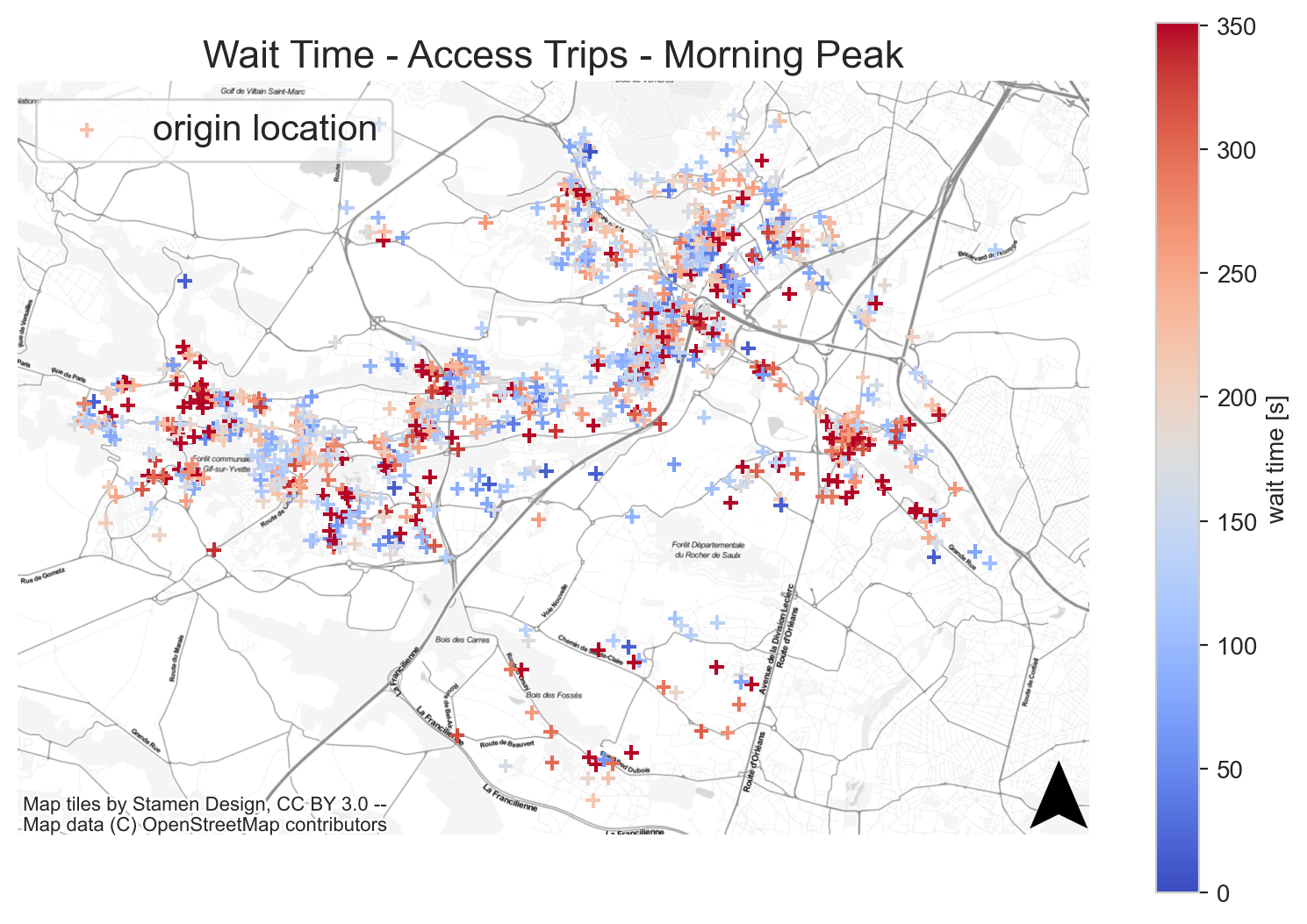}
\caption[Spatial Trend Wait Time]{Spatial Trend Wait Time -  No clear pattern can be identified, indicating low spatial autocorrelation}
\label{map:wt_space}
\end{figure}
\begin{figure}      
\center\includegraphics[width=100mm]{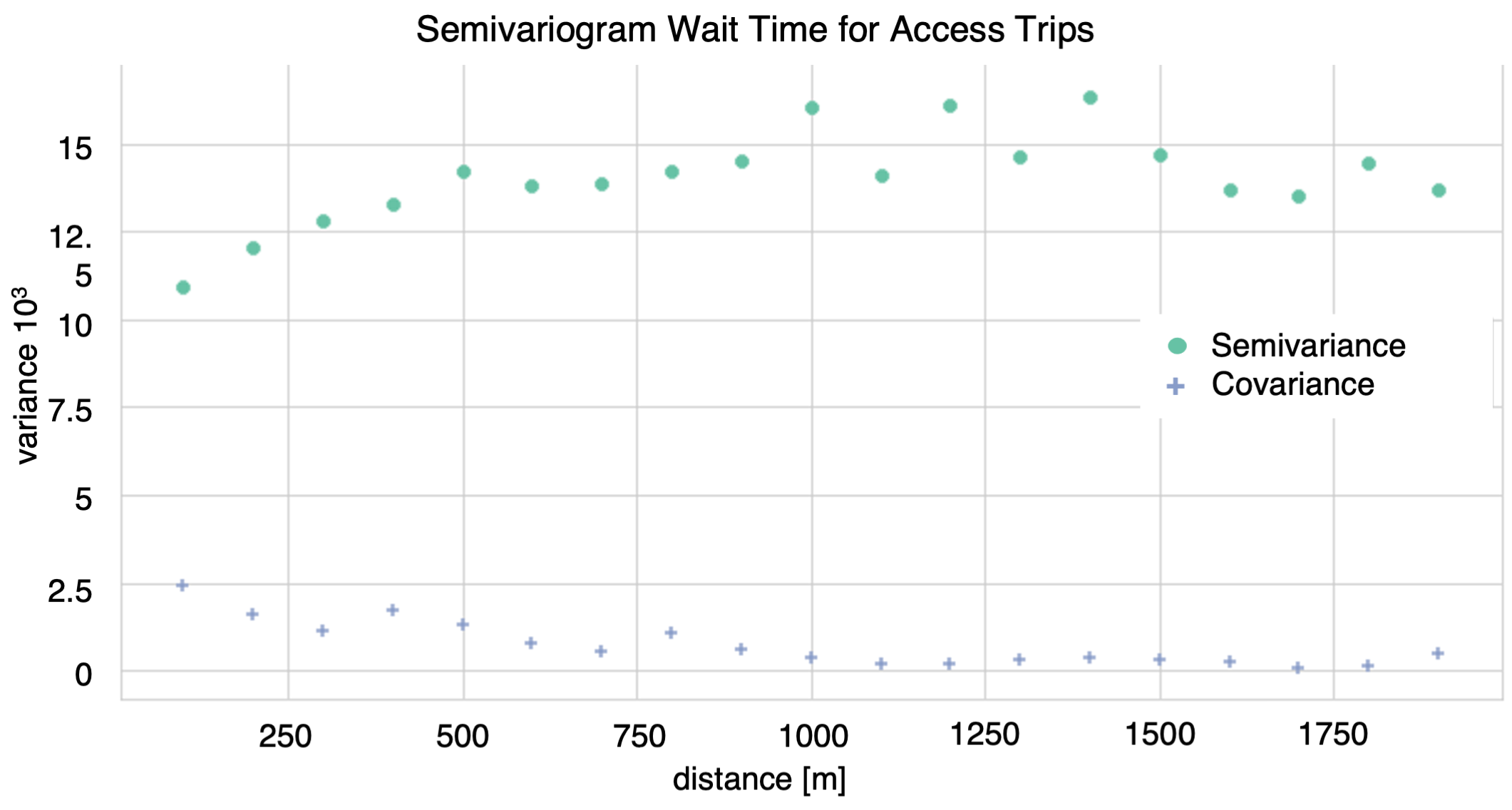}
\caption[Semivariogram Wait Time of Access Trip Observations]{Spatial correlation of wait time observations}
\label{semivar:wt}
\end{figure}
Such trends are not as evident for wait times (Figure~\ref{map:wt_space}) although similarity between observations still decay with distance (Figure~\ref{semivar:wt}).

\begin{figure}
\center\includegraphics[width=148mm]{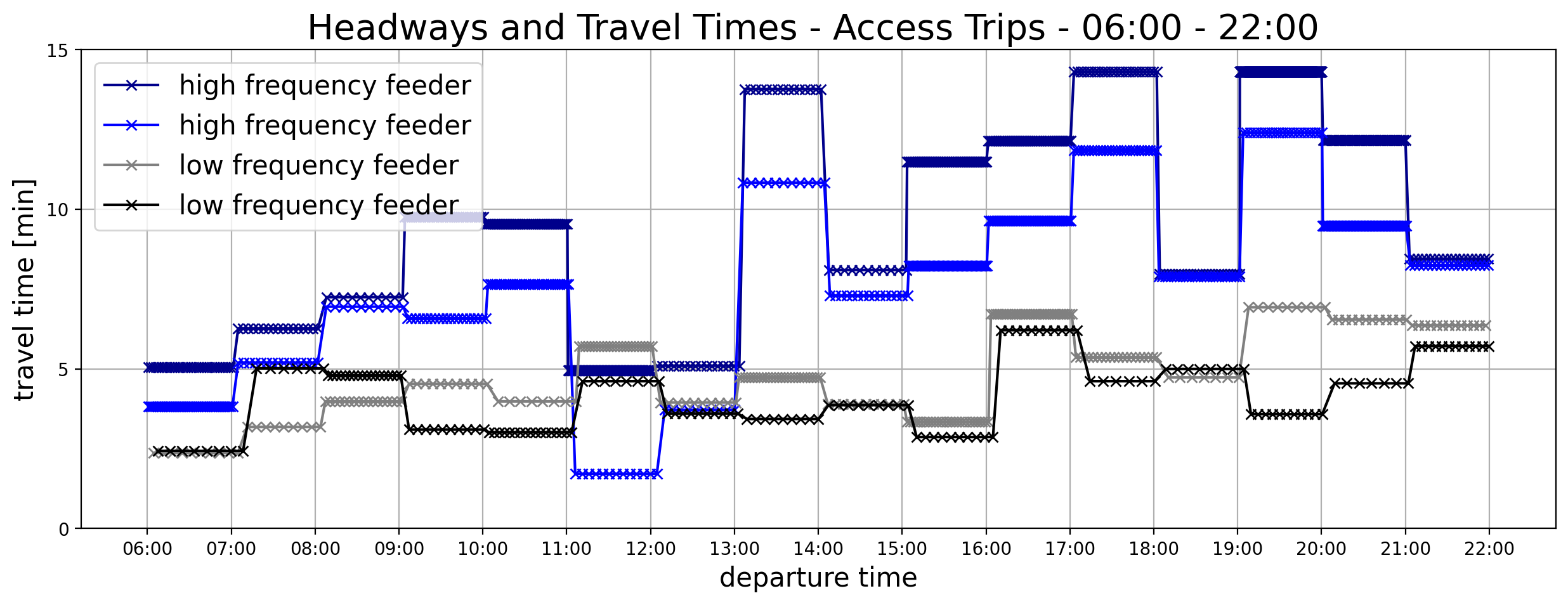}
\caption[Headway and Travel Times in output GTFS]{Headway and Travel Times of some examples of virtual DRT trips. Each departure time of a virtual DRT trip is indicated by a cross. The respective travel time is indicated by the y-axis value.}
\label{fig:headway_results}
\end{figure}

\subsection{Improvement of Accessibility Brought by DRT}
\label{sec:improvement}

Figure~\ref{fig:headway_results} shows headway and travel times of the virtual DRT trips added to the PT graph. We can then compute accessibility on this graph. Note that accessibility varies with the time of day~\eqref{eq:acc}. However, in the following figures we show averages over the time periods mentioned.

\begin{figure}
\noindent\hspace{0.5mm}\includegraphics[width=148mm]{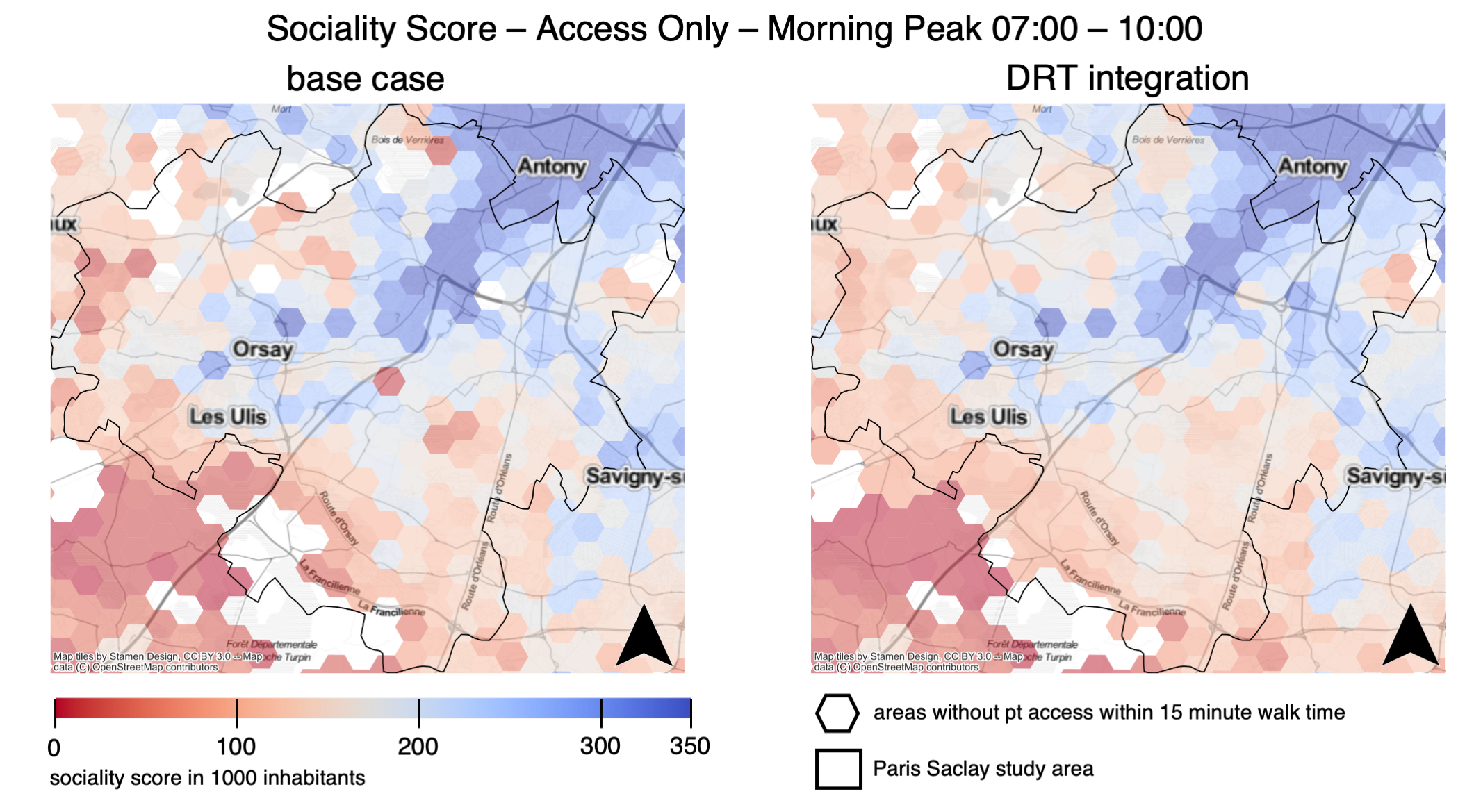}
\caption{Sociality Score - Access Only - Morning Peak 07:00 - 10:00}
\label{map:sc_bc_drt_mp}
\end{figure}

\begin{figure} 
\noindent\hspace{0.5mm}\includegraphics[width=148mm]{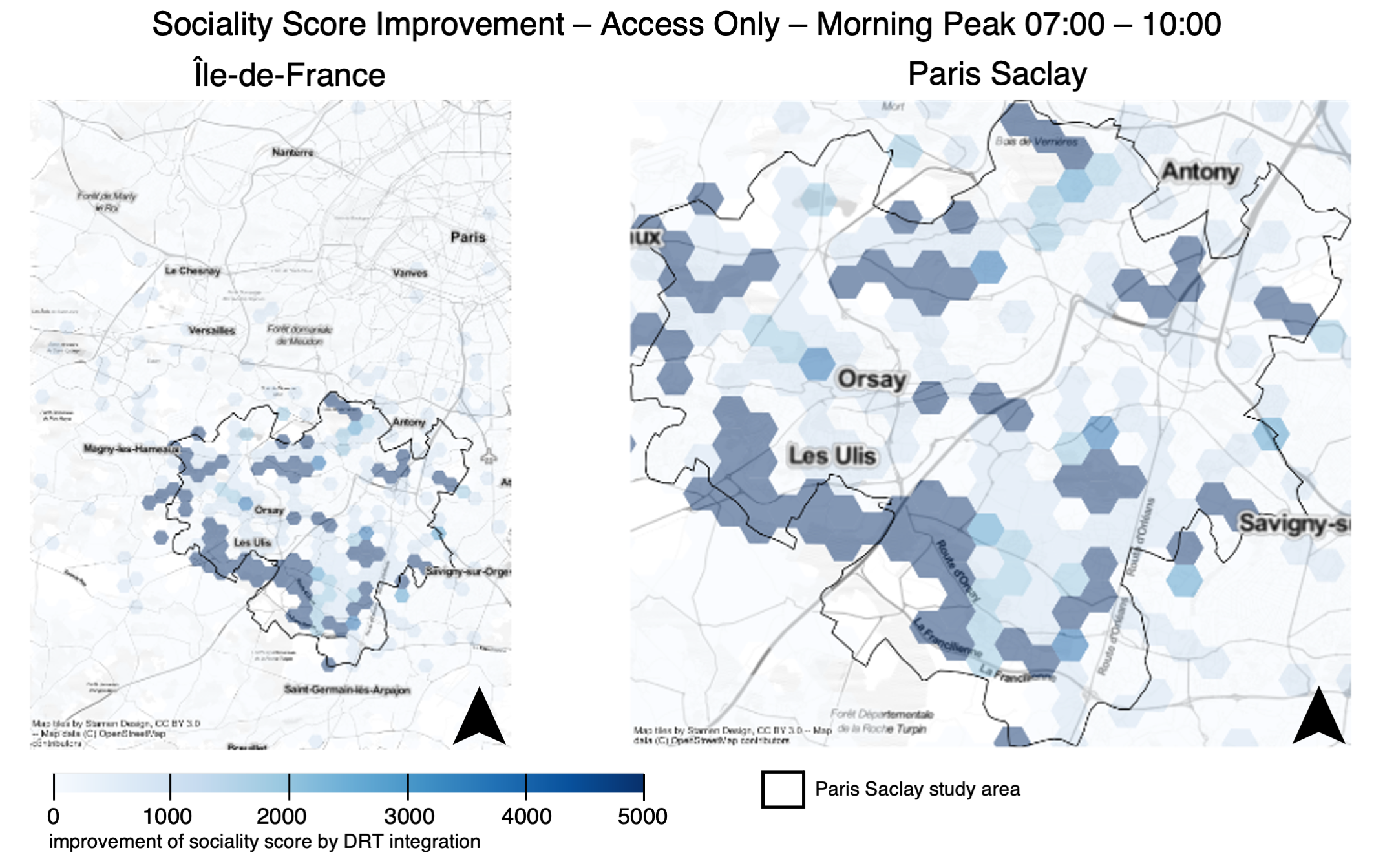}
\caption{Sociality Score Improvement - Access Only - Morning Peak 07:00 - 10:00}
\label{map:sc_diff_bc_drt_mp}
\end{figure}

First, we study a system with DRT access services only (no egress).
Figure~\ref{map:sc_bc_drt_mp} shows that the catchment area is expanded, especially in the south: hexagons with no access to PT within 15 minutes walk, can now use PT. Figure~\ref{map:sc_diff_bc_drt_mp} shows more clearly the improvement in accessibility brought by improved access to PT thanks to DRT. As only access SMS feeder is added in Paris Saclay, the areas outside Saclay do not show any changes, except sligth improvement in some locations, for instant south of Versaille, possibly due to the possibility for travelers starting from there to make changes in Saclay, which are enhanced by DRT.

\begin{figure}
\noindent\hspace{0.5mm}\includegraphics[width=148mm]{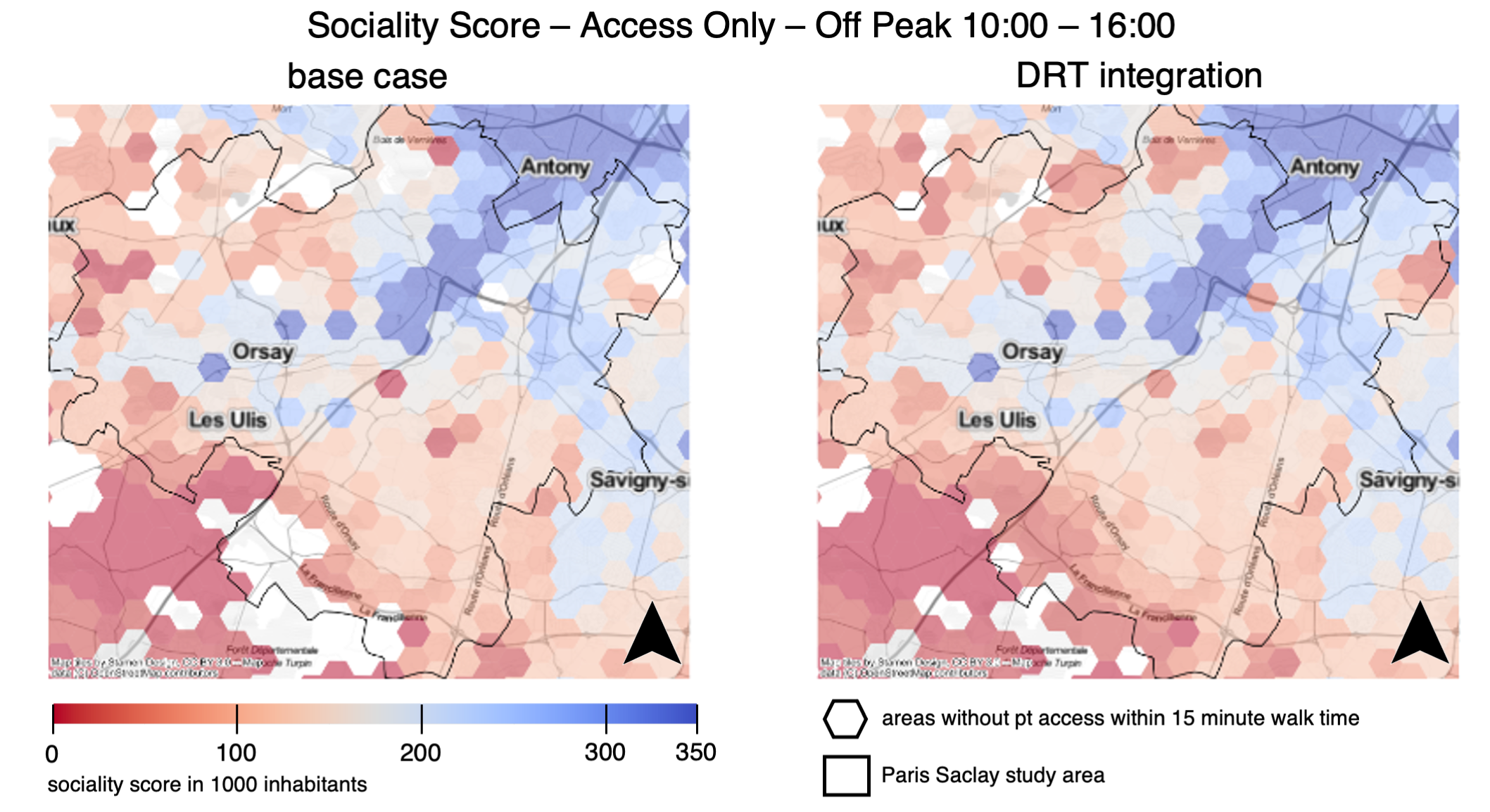}
\caption{Sociality Score - Access Only - Off Peak 10:00 - 16:00}
\label{map:sc_bc_drt_op}
\end{figure}
\begin{figure}
\noindent\hspace{0.5mm}\includegraphics[width=148mm]{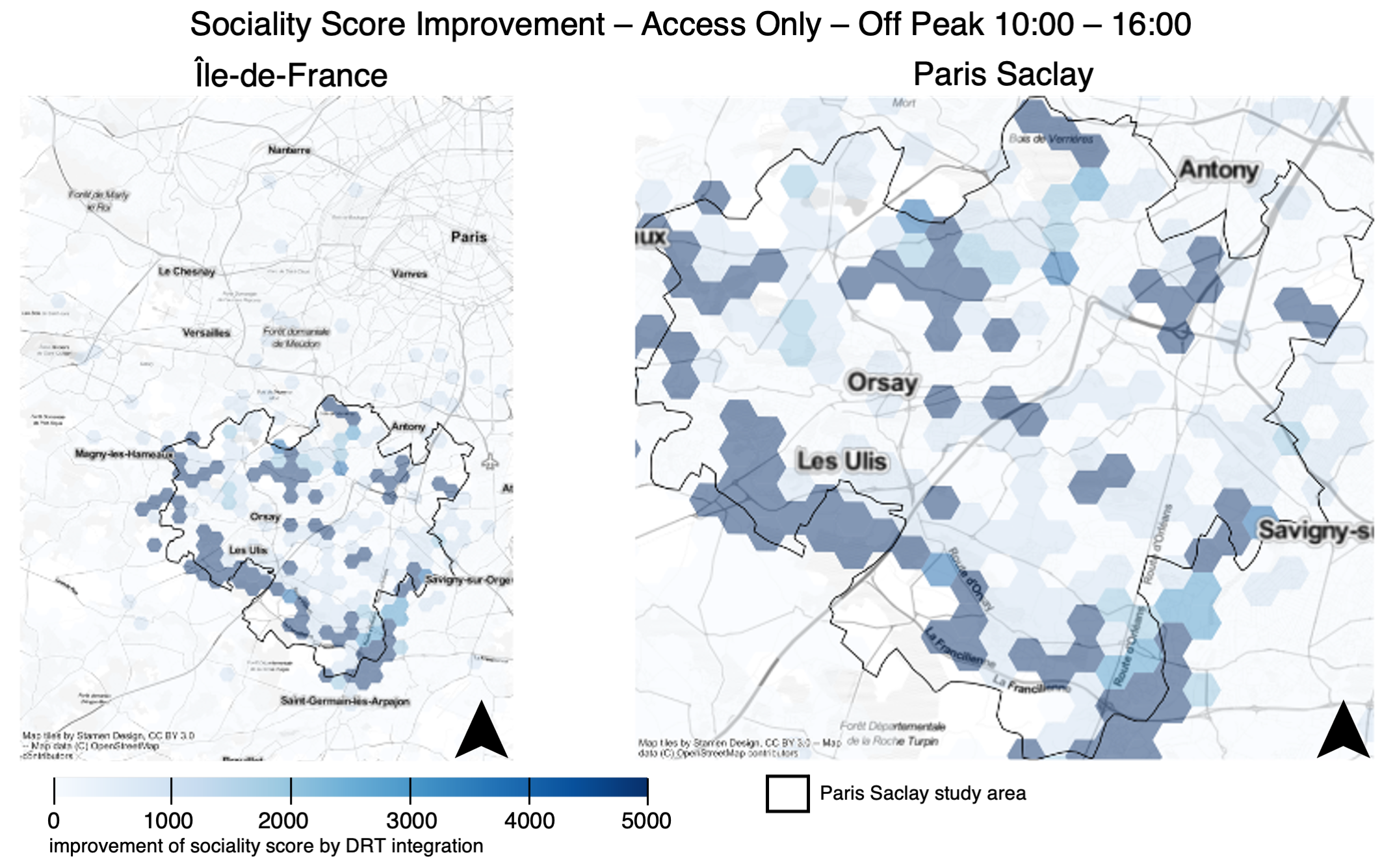}
\caption{Sociality Score Improvement - Access Only - Off Peak 10:00 - 16:00}
\label{map:sc_diff_bc_drt_op}
\end{figure}

Accessibility improvements are even greater in peak hours (Figures~\ref{map:sc_bc_drt_op} and~\ref{map:sc_diff_bc_drt_op}, as DRT compensates for the low frequency of conventional PT.

\begin{figure}
\noindent\hspace{0.5mm}\includegraphics[width=148mm]{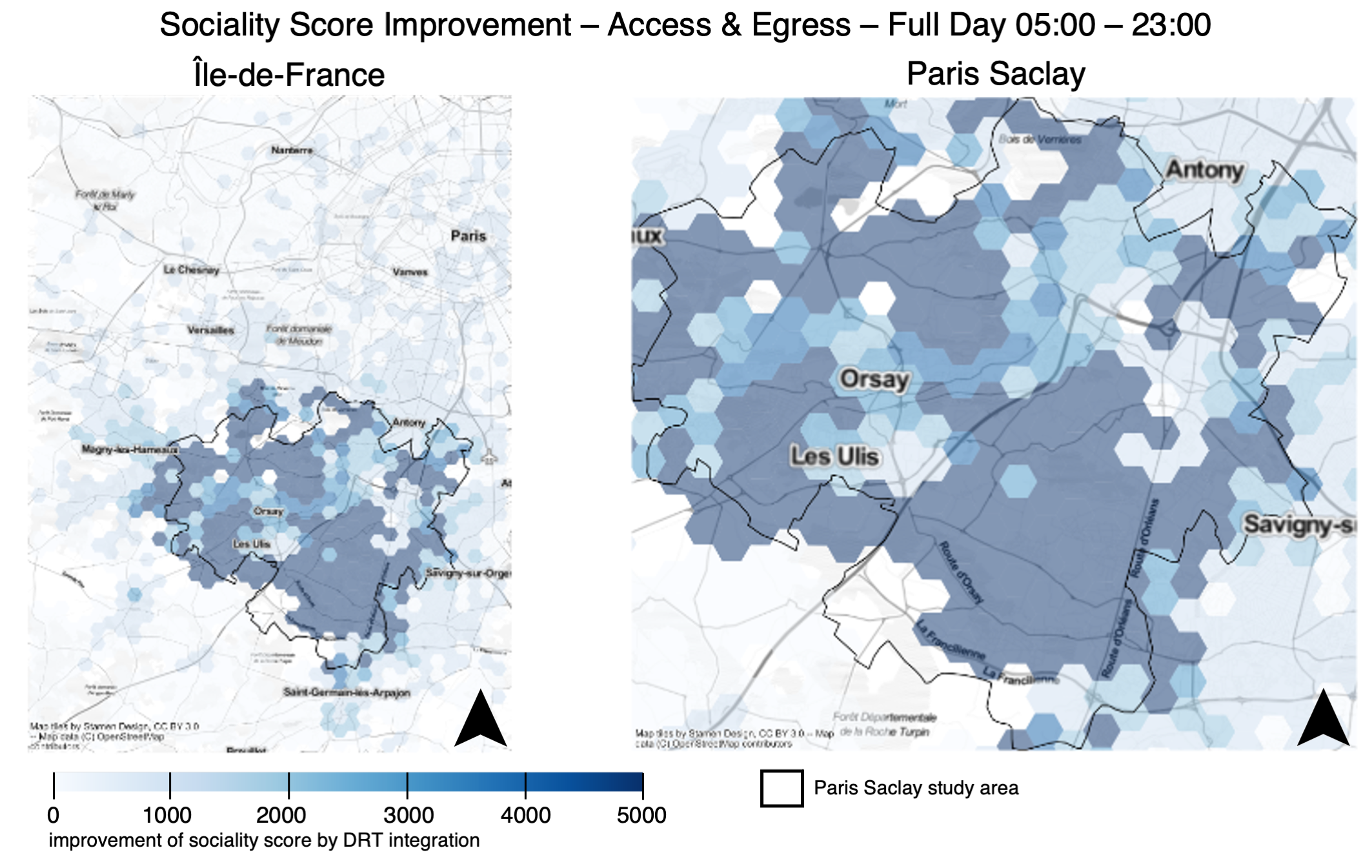}
\caption{Sociality Score Improvement - Access \& Egress - Full Day 05:00 - 23:00}
\label{map:sc_diff_bc_drt_fd}
\end{figure}

Figure~\ref{map:sc_diff_bc_drt_fd} shows the improvement in accessibility when both access and egress trips are added, averaged over the entire day. Improvement is much greater than the access-DRT only case. Moreover, improvement is also visible also outiside Saclay: users from everywhere can now reach opportunities in Saclay faster, thanks to DRT egress connections.

\section{Conclusions}
We proposed a method to compute the impact of SMS on accessibility, based on empirical observations of SMS trips. Our method can support transport agencies and authorities in future deployment of SMS. In our future work, we will empirically validate the results by running simulations where we replaced simulated SMS with our estimated virtual trips. Finally, we will apply our method to car- or bike-sharing feeder and, possibly, on observations from real deployments.

\section*{Acknowledgements}
This work has been supported by The French ANR research project MuTAS (ANR-21-CE22-0025-01) and by BayFrance. It has also been supported by the European Union’s Horizon 2020 research and innovation programme under the Marie Skłodowska-Curie grant agreement no. 899987
Data were provided by the Anthropolis research project at the SystemX Technological Research
Institute, supported by the French government under the “France 2030” program. 
\bibliography{references}

\end{document}